\documentclass[useAMS,usenatbib,usegraphicx]{mn2e}
\usepackage{xspace}

\title{Hydrostatic photoionization models of the Orion Bar}

\author[Y.~Ascasibar et. al.]
{
Y.~Ascasibar, A.~C.~Obreja, and A.~I.~D\'{i}az\\
Departamento de F\'{i}sica Te\'{o}rica, Universidad Aut\'{o}noma de Madrid, Madrid 28049, Spain\\
}

\date{\bf Draft version 2.0 (\today)}

\usepackage{color}
\newcommand{\Referee}[1]{#1}

\newcommand{\dd}{{\rm d}}

\newcommand{\hii}{H{\sc ii}\xspace}
\newcommand{\oric}{$\theta^{1}$ Ori C\xspace}
\newcommand{\sii}{[S{\sc ii}]\xspace}
\newcommand{\siil}{[S{\sc ii}]\,$\lambda\,\lambda$\,6716+6731\xspace}
\newcommand{\ha}{H$\alpha$\xspace}

\begin{document}

\maketitle

\begin{abstract}
Due to its proximity to the Earth and its nearly edge-on geometry, the Orion Bar provides an excellent testbed for detailed models of the structure of \hii regions and the surrounding photon-dominated regions.
In the present study, a self-consistent model of the structure of the Orion Nebula in the vicinity of the Bar is built under the assumption of approximate ionization, thermal, and hydrostatic equilibrium.
It is found that a fairly simple geometry is able to describe the surface brightness profiles of the emission lines tracing the ionized \hii region with a remarkable accuracy, independent of the prescription adopted to set the magnetic field or the population of cosmic rays.
Although we consider different scenarios for these non-thermal components, none of the models is able to provide a fully satisfactory match to the observational data for the atomic layer, and the predicted column densities of several molecular species are always well above the measured abundances.
Contrary to previous studies, we conclude that a more elaborate model is required in order to match all the available data.
\end{abstract}

\begin{keywords}
ISM: HII regions -- ISM: PDR -- ISM: individual (Orion Nebula, M\,42, NGC\,1976) -- ISM: individual (Orion Bar)
\end{keywords}

 \section{Introduction}

\hii regions are extended, low-surface brightness, diffuse nebulae of photoionized gas.
They are associated with regions of ongoing star formation, most commonly found in the disks of spiral galaxies, where young and bright stars provide the copious amount of ionizing ultraviolet (UV) radiation required for such regions to exist.
The spectrum of these nebulae is mainly composed of hydrogen recombination lines and forbidden lines of ions of common elements, superimposed on a weak continuum.
\hii regions can be used to trace star formation from the solar neighborhood to the high-redshift Universe, and their spectrum provides invaluable information about 
the ionizing population of massive stars, as well as the physical conditions of the interstellar medium.

The central source of such a region can be one or several Population I stars of type O or early B, with effective temperatures between 3 and $5 \times 10^{4}$~K, that emit a large number of photons with energies higher than the ionization potentials of hydrogen and helium.
Although these two elements are by far the most abundant ones, metal lines play an important role because they provide the principal cooling mechanism.
At any point in the nebula, the degree of ionization is determined by the equilibrium between electron capture and photoionization.
Ejected photoelectrons carry the excess energy of the photon as kinetic energy, and they contribute through electron-electron and electron-ion collisions to maintaining a Maxwellian velocity distribution with typical electron temperatures between 5000 and 20000 K.

The limit of the \hii region, an ionization front that tends to expand into the surrounding neutral gas at sub-sonic velocities, is usually approximated as a Str\"{o}mgren's sphere with typical radii of the order of parsec.
The \hii region is surrounded by a neutral, photon-dominated region (PDR), where the stellar UV radiation heats the gas and partly dissociates the molecular hydrogen (hence the name photo-dissociation region, which is also often used).
Photons escaping far into the outer molecular region are also absorbed by the dust present in the interstellar medium, which gets thus heated to about 100 K, and re-emitted as an infrared continuum.

Notwithstanding with this simple picture, observations of many \hii regions show signatures of dense neutral condensations scattered in the ionized zone and turbulent motions within the gas with velocities of the order of 10~km~s$^{-1}$.
The geometry of these nebulae is anything but spherical, and theoretical models of their internal structure ought to be constructed on case-by-case basis.

In this work, we attempt to build a self-consistent model of the Orion Bar.
The Orion Nebula (M\,42, NGC\,1976) is part of the Orion Molecular Cloud Complex, at a distance of $437$~pc \citep{Hirota+07}, and it has been studied extensively over the years across the entire electromagnetic spectrum.
Most of the ionizing radiation comes from the star $\theta^1$~Orionis~C, located roughly at the centre of the region.
The Orion Bar, situated at approximately $0.235$~pc from the central star towards the south-east, is a dramatic example of the interface between the ionized gas and the PDR.
Due to its convenient orientation, allowing an almost edge-on perspective, and the rich variety of observations available in the literature, the Orion Bar constitutes an ideal laboratory for testing the PDR physics.

Recent studies \citep{ODe09,Sha09,Pel09} highlight the importance of cosmic rays and magnetic fields in determining the structure of the PDR.
In particular, it has been argued \citep{Pel09} that a relatively strong magnetic field, as well as a population of cosmic rays in equipartition (and thus a density that is much higher than the average galactic background), must be present in the Bar in order to reproduce the observed surface brightness profiles of the H$_{2}$ line at $2.121~\mu$m and the $^{12}$CO$(J = 1-0)$ emission.

The presence of an important magnetic field is supported by polarization observations of the Orion Nebula \citep{Sch98} and the detection of a magnetic field in Orion's Veil with an intensity that is at least an order of magnitude higher than the typical values measured in the cold neutral medium of the Milky Way \citep{Abe06}, but the enhancement of the cosmic ray contribution above the Galactic background is much more poorly constrained from the observational point of view.
A high cosmic-ray density is simply introduced as an additional heating source that can act much deeper into the molecular cloud than the photons from the central star.
As a result, the temperature of the outer regions is considerably higher, providing a better fit to the observed surface brightness profiles.

The present work represents an additional step towards a self-consistent model of the internal structure of the Orion Bar.
As previous studies, it is based on the assumption that the region is in approximate hydrostatic equilibrium: the outward acceleration driven by the radiation field and the pressure of the hot, ionized gas creates an expanding wind that compresses the surrounding medium -- and the magnetic field coupled to it -- until the magnetic pressure is able to halt the process.
Radiative transfer across the different gas phases is solved by the photoionization code Cloudy \citep[last described in][]{Fer98}, including all the relevant processes affecting atoms, molecules, and dust grains, and the predicted line intensities are computed by integrating the physical properties of the gas along the line of sight, following an approach very similar in spirit to \citet{Morisset+05}.

The main improvements with respect to previous work are the inclusion of a detailed treatment of the gravitational acceleration and a more elaborate description of the three-dimensional geometry of the system.
As shown in \citet{Asc10}, the gravitational force (both the mass of the central object as well as the self-gravity of the gas) plays an important role in the outer regions, setting the total extent of the molecular layer and eliminating one free parameter of the model.
In addition, we propose a simple parameterization of the geometry of the \hii region that is able to provide a reasonable fit to the emission line data.
Emission from the atomic layer, though, as well as the column densities of several molecular species, remain difficult to reproduce for any model, with or without gravity.

The details of our photoionization models are thoroughly described in Section~\ref{sec_models}, and their predictions are compared with observational data in Section~\ref{sec_comparison}.
Section~\ref{sec_discussion} is devoted to the physical interpretation of our results, and a brief summary and outlook are provided in Section~\ref{sec_conclusions}.

\section{A simple model of the Orion Bar}
\label{sec_models}

Our model of the Orion Bar is based on the assumption that the gas is in ionization, thermal, and hydrostatic equilibrium at every point.
Under these conditions, the physical properties of a cloud with plane-parallel or spherical symmetry can be efficiently computed with the plasma physics code Cloudy\footnote{All our calculations have been preformed with version C08.00 of the code. The current stable release is publicly available at {http://www.nublado.org}} \citep{Fer98}, a spectral synthesis program designed for the study of low-density environments that are ionized by an external radiation field.
In this section, we describe the parameters of our photoionization models and discuss how the condition of hydrostatic equilibrium and the non-spherical geometry of the Orion Nebula in the vicinity of the Bar may be implemented.

\subsection{Model parameters}
\label{sec_params}

On input, Cloudy requires the user to specify the shape and the intensity of the incident radiation field, the gas density, its chemical composition, and the geometry of the cloud.
For consistency with previous studies, we follow \citet{Pel09} and represent the incident continuum by the sum of the cosmic microwave background, a \citet{Kurucz79} stellar atmosphere with temperature $T=39600$~K and ionizing luminosity $Q(H)=10^{49}$~photons~s$^{-1}$, and a thermal bremsstrahlung component with temperature $T=10^6$~K and luminosity $L=10^{32.6}$~erg~s$^{-1}$ in the $0.5-8$~keV band \citep{Feigelson+05}.
The gas density at the illuminated face is set to $n_0 = 10^{3.2}$~cm$^{-3}$, and a chemical composition appropriate for Orion Nebula \citep{Baldwin+91}, including dust grains and polycyclic aromatic hydrocarbons, is used.

Unless otherwise specified, Cloudy assumes that the gas density stays constant over the cloud, but many density and pressure laws can be used instead.
We model the Orion Bar by enforcing hydrostatic equilibrium with the \emph{constant pressure} command.
Contrary to what the name suggests, this option does \emph{not} keep the total pressure fixed, but adjusts its value so that the total acceleration vanishes at every point,
\begin{equation}
\frac{1}{\rho (r)} \frac{dP(r)}{dr} = a(r)
\end{equation} 
and therefore
\begin{equation}
P(r) = P_0 + \int_{0}^{r}\rho(x)\,a(x)\ dx
\label{eq_P}
\end{equation}
where $P_0$ denotes the initial pressure at the illuminated face, and the acceleration to be balanced includes the terms due to the absorption of photons (pushing the gas away from the central star) and the gravitational force that pulls the whole cloud towards the centre.
The latter is included by means of the \emph{gravity} command, described in \citet{Asc10}.
The gravitational acceleration is calculated as $g(r) = - 4 \pi G M(r)/ r^{2}$, assuming spherical symmetry, and the mass inside radius $r$ includes the contributions of both the gas and the stars in the cloud.
The distribution of the gas mass is computed self-consistently by the code, $ M_{\rm gas}(r) = \int_0^r \rho (x) 4 \pi x^{2} \mathrm{d}x $, and the \oric multiple system is modeled as a point mass of 50~M$_{\odot}$ \citep{Kra07} located at the centre of the region.

The total pressure
\begin{equation}
P(r) = P_{\rm gas} + P_{\rm ram} + P_{\rm turb} + P_{\rm mag} + P_{\rm lines}
\end{equation}
that appears in equation~(\ref{eq_P}) is the sum of the thermal pressure of the gas $P_{\rm gas} = nkT$, the terms $P_{\rm ram} = \rho v_{\rm wind}^{2}$ and $P_{\rm turb} = \rho v_{\rm turb}^{2} /2$ induced by the uniform and turbulent motions, respectively, the magnetic pressure $P_{\rm mag} = B^{2} /8\pi$, and the contribution $P_{\rm lines}$ \citep{FerlandElitzur84,ElitzurFerland86} of the trapped emission lines.
Although winds are not considered in our model, it includes a small turbulent velocity field of 2~km~s$^{-1}$.
Regarding the magnetic field, we consider the same scenarios as \citet{Pel09}.
In the \emph{gas pressure} model, only the thermal pressure of the gas and the turbulent pressure terms are taken into account.
The \emph{magnetic pressure} model adds a tangled magnetic field whose intensity at the illuminated face of the cloud is $B_{0}=8~\mu$G, and its equation of state is given by $\gamma =2$ (i.e. $B/B_{0} = n / n_{0}$).
In both scenarios, the density of cosmic rays is constant and equal to the average galactic background.
The \emph{enhanced cosmic rays} model assumes the cosmic-ray density to be in equipartition with the magnetic field, resulting in a much higher abundance of relativistic particles.

For each scenario, we ran a grid of Cloudy models with initial distances from \oric ranging from $0.05763$ to $0.5763$~pc in logarithmic steps of $0.01$.
A sample Cloudy script used in the preparation of our model grids is shown in Appendix~\ref{app_script}.
Each model saves the emissivities of two emission lines from the ionized region (\siil and \ha) and one emission line produced in the atomic layer (H$_2\,2.121~\mu$m), the extinction towards the end of the cloud, and the volume densities of six molecular species (CO$^+$, SO, CN, CS, SO$^+$, and SiO), all of them as a function of depth into the cloud (i.e. the distance from the illuminated face).

\subsection{Geometry}

One of the aims of the present work is to show that the overall geometry of the Orion Nebula plays an important role on its physical conditions and observable properties, which can be exquisitely probed in the region near the Bar due to its privileged orientation.
Previous studies have constrained the geometry of the cloud using the surface brightness profile of the emission lines associated to the \hii region.
More specifically, the \siil line displays a sharp peak at the interface between the ionized and the neutral layers, thus providing an excellent tracer of the position of the ionization front \citep[see e.g.][]{Baldwin+91,Wen95}.

\begin{figure}
\centering\includegraphics[width=8cm]{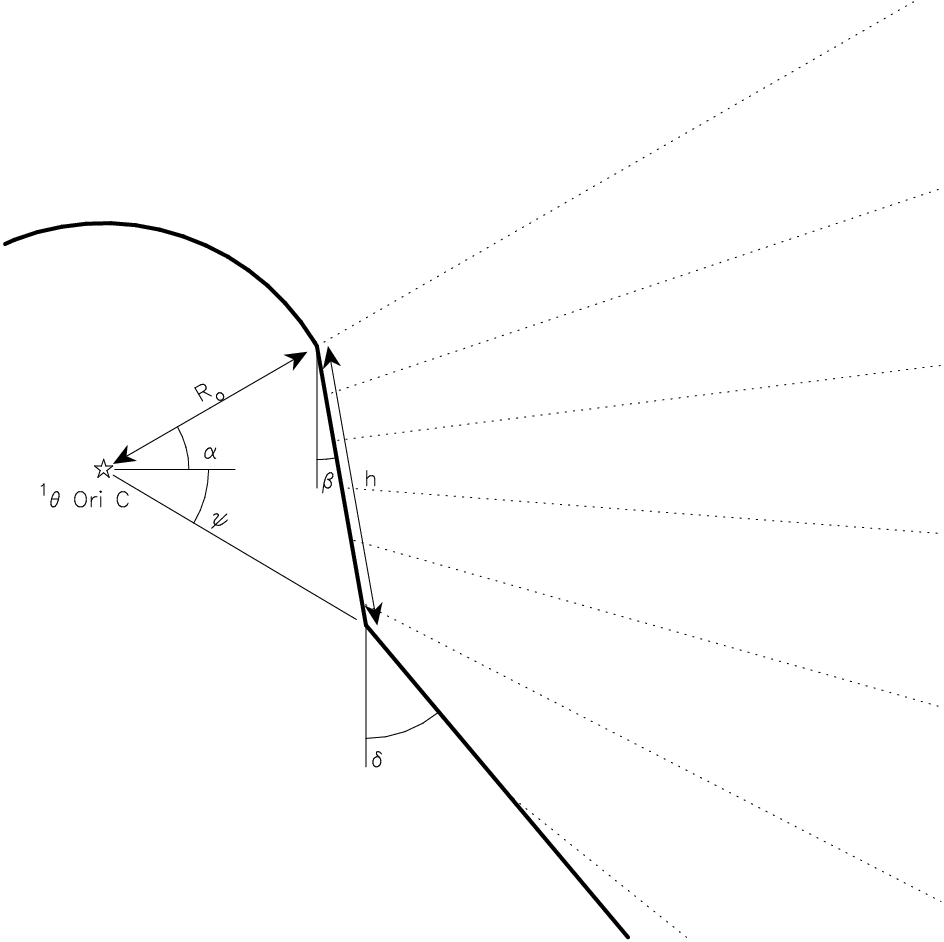}
\caption
{
Geometry of the illuminated face of the cloud, described by the parameters $R_0$, $\alpha$, $\beta$, $\delta$, and $h$ (see text for a detailed explanation).
The $x$- and $y$-axes represent the offset with respect to \oric and the coordinate along the line of sight, respectively.
}
\label{fig_geometry}
\end{figure}

In the vicinity of the Bar, the projected distance between the maximum of the observed emission and the central star is 111~arcsec (0.235 pc, assuming a distance of 437 pc).
Approximating the Bar as a plane-parallel slab of thickness $h$, located at a distance $R_0$ from \oric with an inclination angle $\beta$ with respect to the line of sight, the data suggest the values $h \sim 0.115$~pc, $R_0 \sim 0.114$~pc, and $\beta \sim 7^\circ$ \citep{Pel09}.

We consider a very similar layout, depicted in Figure~\ref{fig_geometry}, that also includes the contribution of the adjacent regions of the Orion Nebula.
The interior region, towards the central star, is approximated as a sphere of radius $R_0$ centered at the location of \oric, which we take as the origin of coordinates.
The transition between the spherical region and the Orion Bar takes place at an angle $\alpha$ with respect to the plane of the sky, which is an additional free parameter of our model.
The Bar itself, as well as the outer region, are modeled as straight lines whose angles with respect to the line of sight are denoted by the parameters $\beta$ and $\delta$, respectively.
The length of the Bar region is specified by the parameter $h$, whereas the outer region is assumed to continue well beyond the observed field.

The region interior to this curve is empty, or, more precisely, filled by a hot, tenuous gas that can be detected in X-rays \citep{Guedel+08} and provides thermal pressure support for the warm ($T \le 10^4$~K) dense gas that is responsible for the optical and infrared emission in which we are interested.
In Cloudy parlance, the curve depicted in Figure~\ref{fig_geometry} defines the illuminated face of the cloud.
For the outer region to receive direct illumination from \oric, as assumed by our models, it must not be behind the shadow cast by the Bar (angle $\psi$ in Figure~\ref{fig_geometry}).
Thus, we chose the parameterization
\begin{equation}
 \delta = \eta\, \left( \frac{\pi}{2} + \psi \right) + \left( 1 - \eta \right)\, \beta
\label{eq_eta}
\end{equation}
where $0<\eta<1$.

This geometry is implemented as a summation over a grid of Cloudy models with spherical symmetry and different distances between the illuminated face of the cloud and the central star (light dashed lines in Figure~\ref{fig_geometry} illustrate the separation between models).
The spherical region corresponds to a single Cloudy model, fully characterized by the value of $R_0$, while both the Bar and the outer region involve many models each.

The column density of a given species $i$ is given by the integral along the line of sight
\begin{equation}
N_{i}(x) = \int_{-\infty}^{\infty} n_{i}(x,y)\ \dd y
\end{equation}
where the volume density $n_i(x,y)$ is evaluated from the output of the Cloudy model that is appropriate for each position.
Surface brightness profiles of the emission lines have been computed as
\begin{equation}
S_{i}(x) = \int_{-\infty}^{\infty} \frac{ \varepsilon_{i}(x,y) }{ 4\pi }\ 10^{-0.4 A_i(x,y)}\ \dd y
\end{equation} 
where $\epsilon_i(x,y)$ denotes the emissivity per unit volume of the line, and the amount of dust extinction
\begin{equation}
A_{i}(x,y) = \int_{-\infty}^{y} \frac{ \partial A_i }{ \partial y' }(x,y')\ \dd y'
\end{equation}
includes all the foreground material between any given point and the observer.
The differential extinction $\frac{ \partial A_i }{ \partial y' }(x,y')$ is estimated from the radial increment of the total extinction in the V band output by Cloudy and then converted to the rest wavelength of the line.

 \section{Comparison with observations}
\label{sec_comparison}

\subsection{\hii region}

As will be shown below, the observational properties of the ionized region (more precisely, the surface brightness profiles of its emission lines) are not very sensitive to the details of the magnetic field or the prescription adopted to establish the population of cosmic rays.
The exact position of the ionization front depends on a combination of the gas density at the illuminated face, the intensity of the ionizing radiation from \oric, and the distance to the Orion Nebula.
Once these (degenerate) parameters are specified, the shape of the emission profiles is entirely determined by the geometrical configuration of the system.

In our model, this geometry (the distance from the central star to the illuminated face) is described by the values of five free parameters: the radius of the inner region $R_0$, the transition to the Bar $\alpha$, its inclination with respect to the line of sight $\beta$, its length $h$, and the angle $\delta$ -- or, equivalently, the parameter $\eta$ defined in expression~(\ref{eq_eta}) -- that defines the orientation of the outer region with respect to the line of sight.
We estimate the values of these parameters by fitting the observed surface brightness profiles of the \siil\AA\ and \ha emission lines.

The surface brightness profile of the \siil line across the bar has been obtained from two narrow-band images, taken by \citet{Pel09} with the Southern Astrophysical Research Telescope, centered at $\lambda=6723$ and $6850$~\AA\ with bandwidths of 45 and 95~\AA, respectively.
For the H$\alpha$ line, we use the observations of \citet{Wen95}.
Both data sets were constructed as continuum-subtracted averages over $20$~arcsec-wide swathes, using similar cuts.
Since these data have already been corrected for dust extinction, we set $A_i=0$ when modelling both emission lines.

The best-fitting values of $R_0$, $\alpha$, $\beta$, $\eta$, and $h$ have been found by means of the FiEstAS sampling technique \citep{Ascasibar08}, a Monte Carlo integration scheme based on the Field Estimator for Arbitrary Spaces \citep[FiEstAS;][]{AscasibarBinney05,Ascasibar10}.
In order to quantify the quality of the fit to the observational data, we compute the reduced $\chi^{2}$ as
\begin{equation}
\chi^{2} = \frac{ 1 }{ N_{\rm obs} } \sum_{ j=1 }^{ N_{\rm obs} }
\frac{ \left[\, S_{\rm obs}(x_{j}) - S_{\rm model}(x_{j}) \,\right]^2 }{ \sigma_j^{2} }
\label{eq_chi2}
\end{equation}
where $S_{\rm obs}(x_{j})$ denotes the observations at a projected distance $x_{j}$ from the central star, $S_{\rm model}(x_{j})$ are the corresponding model predictions, and the sum over the index $j$ corresponds to the $N_{\rm obs}$ observational data points, whose errors have been characterized by a standard deviation $\sigma_j = 5 \times 10^{-14}$ and $5 \times 10^{-13}$~erg~s$^{-1}$~cm$^{-2}$~arcsec$^{-2}$ for the \siil and the \ha lines, respectively.

\begin{figure}
\centering\includegraphics[width=8cm]{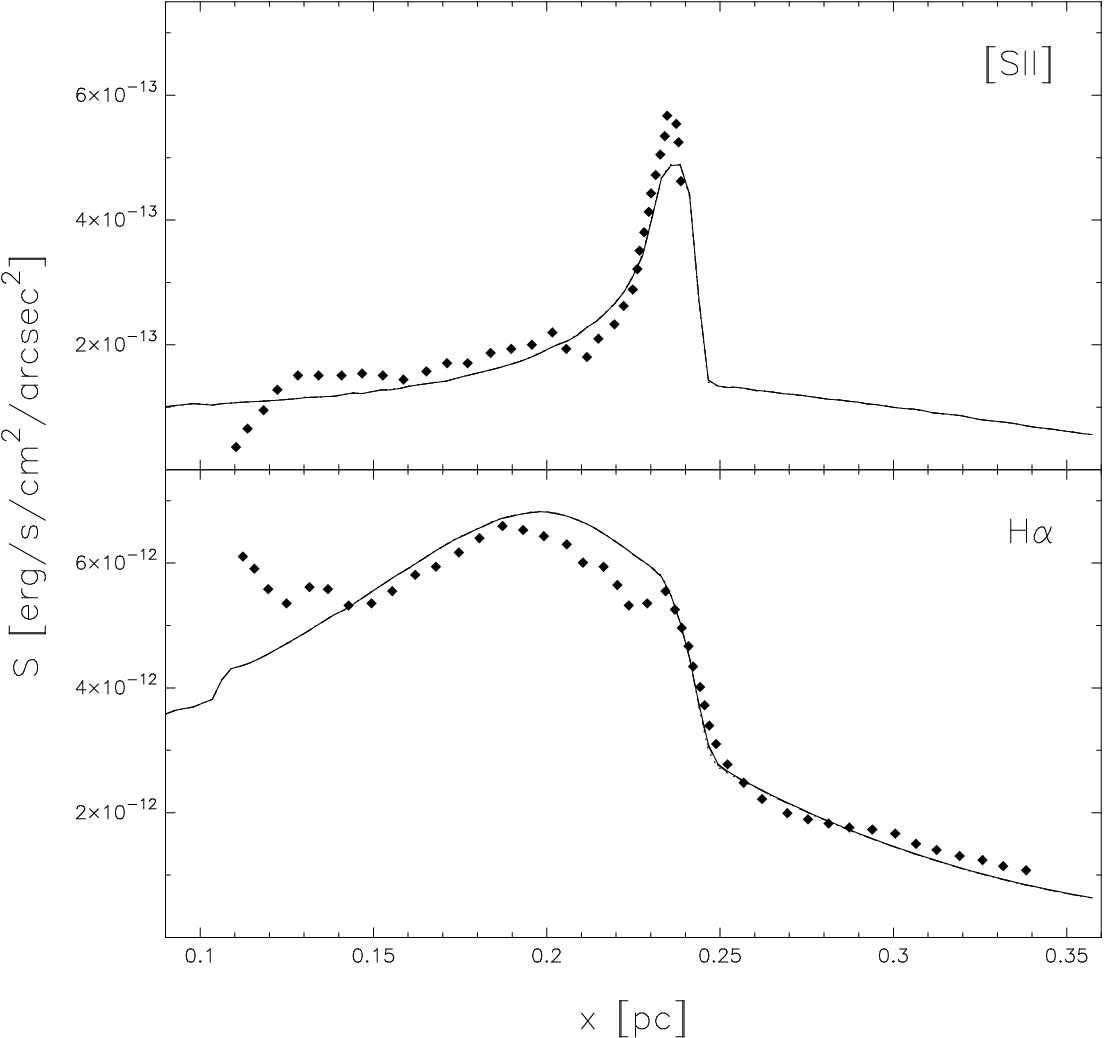}
\caption
{
Observed surface brightness profiles of the \siil and \ha emission lines (dots), compared to the theoretical model predictions.
The gas pressure, magnetic pressure, and enhanced cosmic rays models have been plotted as dotted, dashed, and solid lines, respectively, but they lie virtually on top of each other.
}
\label{fig_ionized}
\end{figure}

The results are shown in Figure~\ref{fig_ionized}, where the surface brightness profiles obtained for $R_0 = 0.122$~pc, $\alpha = 28.3\,^\circ$, $\beta = 2.3\,^\circ$, $h = 0.05$~pc, and $\delta = 45.15\,^\circ$ are compared with the observational data points.
Our three scenarios for the magnetic field and cosmic rays yield exactly the same prediction for the emissivities within the ionized region of the nebula.
Although the value of the best-fit $\chi^2 = 0.62$ is somewhat arbitrary, since it depends on the adopted $\sigma_i$, we judge from Figure~\ref{fig_ionized} that the proposed geometry is able to provide a reasonable description of the Orion Bar up to the ionization front, regardless of the assumptions made concerning the non-thermal components.

\subsection{Atomic region}

In order to test the ability of our models to describe the structure and physical properties of the atomic layer, we will now focus on the H$_2$ S$(1-0)$ transition at $2.121~\mu$m, comparing the model predictions with the observational data reported in \Referee{\citet{Wer96}}.
The spatial cut from which the H$_2$ data were obtained is different from that of the [S{\sc ii}] and \ha lines.
Although the [S{\sc ii}] profile shows little variation \citep[see e.g.][]{Henney+05a}, it has been shown \Referee{\citep{Wer96,YOw00,Walmsley+00,All05}} that the surface brightness of the H$_2$ line may change considerably when using different cuts perpendicular to the Bar, which introduces some uncertainty in the comparison.
\Referee
{
Moreover, the data represent a summation parallel to the bar, whereas the H$_2$ emission is actually concentrated in rather thin filaments with typical widths of $\sim 8$ arcsec, about 0.016 pc.
}

Another potential issue is the estimation of the amount of dust extinction in the models.
\Referee
{
On the one hand, our models may overestimate the value of $A_{\rm V}$, perhaps due to the assumed cross-section for dust attenuation \citep{All05} and/or an excessive amount of dust in the atomic region.
On the other hand}, the conversion between the visual extinction $A_{\rm V}$ returned by Cloudy and the extinction at the infrared wavelength $\lambda = 2.121~\mu$m of the H$_2$ emission line
\begin{equation}
 A_{2.121\,\mu{\rm m}}\ \approx\ 0.14\, A_{\rm V} 
\label{eq_extinction}
\end{equation}
may be obtained by applying a \citet{Cardelli+89} reddening curve with $R=5.5$, in accordance with observations of the Orion Nebula.
This estimate is based on the observed reddening curve of stellar spectra, where the extinction includes both the absorption and scattering of light by the dust grains.
For an extended source, like the Orion Nebula, a large fraction of the photons (those that are scattered a small angle away from the light of sight) will be compensated by similar small-angle events affecting nearby rays, and the effective scattering opacity
\begin{equation}
 \sigma_{\rm scat} = \sigma_{\rm s}\ ( 1 - g ) 
\end{equation}
will be given by the product of the total scattering cross-section $\sigma_{\rm s}$ and the grain asymmetry factor $g$, defined as the average $\langle \cos\theta \rangle$ over the angle $\theta$ between the incident and the scattered photon.
Since the wavelength dependence of the asymmetry factor is determined by the chemical composition, shape, and size distribution of the dust grains, and these variables may vary with the spatial location within the nebula (they depend, for instance, on the gas density and the high-energy radiation field), there is \Referee{some} uncertainty associated to equation~(\ref{eq_extinction}).
\Referee
{
Nevertheless, an independent estimation of the nebular extinction in the Orion Nebula based on the hydrogen recombination lines from the Balmer, Paschen, and Brackett series \citep{Bautista+95} is compatible with $A_{2.121\,\mu{\rm m}} = 0.14\, A_{\rm V}$, suggesting that the effects of scattering are not very important in this case.
}

\begin{figure}
\centering\includegraphics[width=8cm]{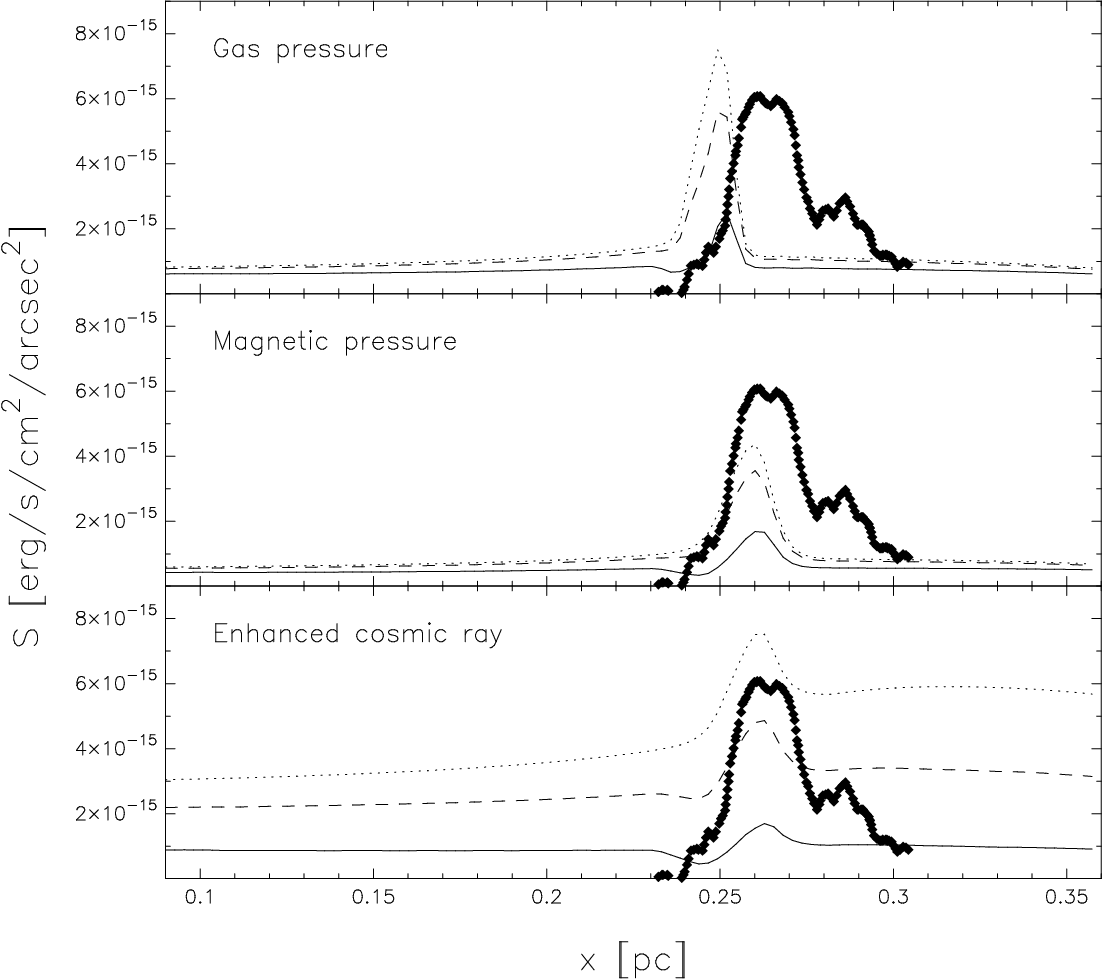}
\caption
{
Surface brightness profiles of the H$_2\,2.121~\mu$m emission line predicted by the gas pressure (top), magnetic pressure (middle), and enhanced cosmic rays (bottom) models.
Dotted lines show the total, unextincted emission. Dashed and solid lines \Referee{correspond to $A_{2.121\,\mu{\rm m}} = 0.028\, A_{\rm V}$} and $A_{2.121\,\mu{\rm m}} = 0.14\, A_{\rm V}$, respectively.
}
\label{fig_H2_line}
\end{figure}

The predicted surface brightness profiles of the H$_2\,2.121~\mu$m line for the gas pressure, magnetic pressure, and enhanced cosmic rays models are shown on the different panels in Figure~\ref{fig_H2_line}.
It is worth noting that, while the emission lines of the ionized region have been used to constrain the free parameters that set the geometry of the illuminated face, the surface brightness profile of the H$_2$ line is a genuine prediction of the models.

Our results suggest that none of the scenarios we have considered is able to reproduce the observed emission at the quantitative level.
\Referee
{
In particular, all models underestimate the maximum intensity of the H$_2$ line by more than a factor of three.
We have verified that the discrepancy between models and data is not a consequence of the prescription adopted for dust extinction.
In the gas pressure scenario, the separation between the peaks of the \sii and the H$_2$ emission is always too small, whereas the magnetic pressure model fails to explain the observed surface brightness even for $A_{2.121\,\mu{\rm m}} = 0$.
The enhanced cosmic rays model seems slightly more promising, but the precise amount of dust extinction is critical.
In addition, the density gradient in the molecular region should be much steeper than predicted by our gravitational models in order to reproduce the observed lack of emission outside the main peak.
}

\subsection{Molecular region}
\begin{figure*}
\centering\includegraphics[width=15cm]{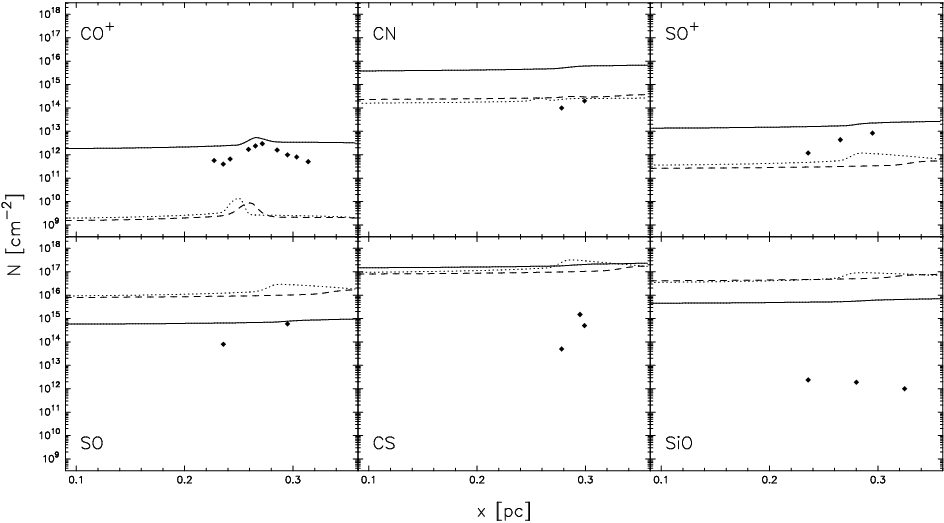}
\caption
{
Observed column densities of CO$^{+}$ \citep{Sto95,Fue03}, CN \citep{Sim97}, SO$^{+}$ \citep{Fue03}, SO \citep{Jan95}, CS \citep{Sim97,Hog95}, and SiO \citep{Sch01}, compared to the predictions of the gas pressure (dotted), magnetic pressure (dashed), and enhanced cosmic rays (solid lines) models.
}
\label{fig_molecules}
\end{figure*}

The abundances of different molecular species provide a direct probe of the extent and the physical conditions of the outer, molecular layer.
As in \citet{Pel09}, we compare the column density profiles predicted by the models with observations of CO$^{+}$ \citep{Sto95,Fue03}, CN \citep{Sim97}, SO$^{+}$ \citep{Fue03}, SO \citep{Jan95}, CS \citep{Sim97,Hog95}, and SiO \citep{Sch01}.

Observed column densities should, however, be treated with some caution, given the many assumptions involved in their derivation.
For instance, \citet{Fue03} assume a constant rotational temperature of 10~K for all the molecular levels, and all lines are considered to be optically thin.
For the CN and CS column densities, \citet{Sim97} combined their observations with previous data from \citet{Fue96} and \citet{Wer96}.
They assumed kinetic temperatures of $40 - 100$~K and used CS rates for the unknown CN collisional rate coefficients.
CN optical depths were estimated by an escape probability code, assuming a single component.
However, the CS column density of \citet{Hog95} was derived using statistical equilibrium calculations that assumed dense clumps to be embedded in a lower density medium, and the same type of analysis was employed by \citet{Jan95} to infer the column density of SO.
The SiO column densities of \citet{Sch01} were obtained from observations of three transitions, assuming the same temperature in the molecular region, $85 \pm 30$~K, as \citet{Hog95}.

For this reason, as well as several factors that may affect the accuracy of the models (see the discussion below), the comparison between model predictions and observational data is far from straightforward.
With this caveat in mind, our results for the six molecular species considered are shown in Figure~\ref{fig_molecules}.

The CO$^{+}$ column density, shown on the upper left panel, is reasonably well reproduced by the enhanced cosmic rays model.
The intensity of the peak, as well as the offset at which it occurs, are consistent within the uncertainties, and the model also explains the extended wings of the column density profile.
However, the gas pressure and magnetic pressure models predict about a factor of $~100$ smaller column densities than observed, in agreement with the findings of \citet{Pel09}.
On the contrary, for the CN molecule, represented in the top middle panel, these models are remarkably close to the observational data of \cite{Sim97}, while the enhanced cosmic rays predictions are now two orders of magnitude above the observed values.

In both cases, the shape of the profiles is arguably a little too flat.
Such a slow increase of the column density when moving away from the central star into the molecular region is even more evident for SO$^{+}$ (shown on the top right panel), whose abundance is bracketed by the predictions of the gas/magnetic pressure and enhanced cosmic rays models, and the SO molecule, plotted on the bottom left panel, where the models with Galactic cosmic rays overestimate the observed column density, and none of them reproduces the observed gradient.
The same occurs for CS on the bottom middle panel, where all models are more than two orders of magnitude above the data \citep[in contrast with the findings of][where the magnetic model gives a good match]{Pel09}, and SiO, on the bottom right panel, where the discrepancy reaches four orders of magnitude.

 \section{Discussion}
\label{sec_discussion}

To summarize, our models of the Orion Bar seem to provide an excellent fit to the \hii region, but they do a much poorer job for the atomic and molecular layers.
On the one hand, they tend to overestimate the molecular column densities under all the scenarios considered for the intensity of the magnetic field and the population of cosmic rays.
On the other hand, the comparison between the predicted and observed intensities of the H$_2$ emission line hints that the models also overestimate the amount of dust in the \Referee{atomic} layer.

One possible reason is, of course, that our models are too simple to describe all the relevant features of the Orion Nebula.
Given the computational cost of building one of our model grids, a full exploration of the parameter space (where the cost increases exponentially with the number of free parameters) is well beyond the scope of the present work.
Nevertheless, we have run several single Cloudy models to test whether a better agreement with the observations could be achieved by varying the density of the gas at the illuminated face, the intensity and equation of state of the magnetic field, or the amount of mass in stars.
\Referee{Alternatively}, it is also possible that the proposed models are incomplete, not because they lack the required flexibility, but because they do not include all the relevant physics.
\Referee{We briefly discuss the validity of some of the main assumptions of the model -- most notably, the approximation of hydrostatic equilibrium.}

\subsection{\Referee{Gas density and magnetic field}}

To a certain extent, the initial gas density of the models can be thought of as a global scaling factor.
Roughly speaking, column densities scale approximately proportional to $n_0$, emissivities increase as $n_0^2$, and distances are proportional to $n_0^{-1}$.
Although there is some freedom in the choice of the optimal value of this parameter, especially if one allows for errors in the estimation of the distance to the Orion Nebula, the observed line ratios of several emission lines impose severe constraints on the ionized gas density.
Moreover, once the geometry is adjusted to fit the surface brightness profiles in the \hii region, the predictions for the other quantities are remarkably insensitive to the precise value of $n_0$.

\begin{figure}
\centering\includegraphics[width=8cm]{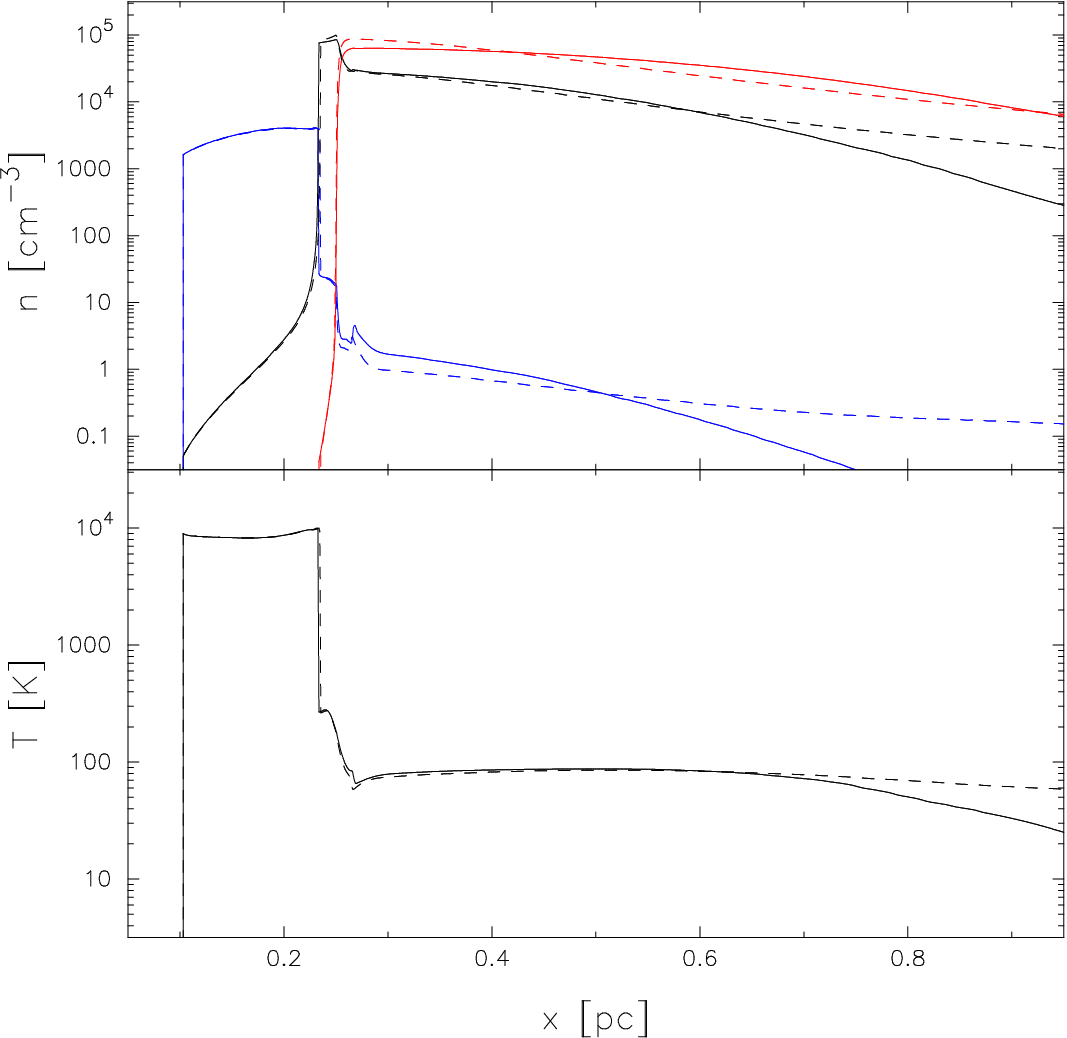}
\caption
{
\Referee{
Number density of ionized, atomic and molecular hydrogen (shown as blue, black, and red lines on the top panel, respectively), as well as electron temperature (bottom panel) for the enhanced cosmic rays case (solid lines) and a model with $B_0 = 50~\mu$G and $\gamma = 1$ (dashed lines).
}
}
\label{fig_magnetic}
\end{figure}

As explained in Section~\ref{sec_params}, the magnetic field is described by the equation of state
\begin{equation}
 \frac{B}{B_0} = \left( \frac{n}{n_0} \right)^{\gamma/2}
\end{equation}
with $B_0 = 8~\mu$G, $n_0 = 10^{3.2}$~cm$^{-3}$, and $\gamma = 2$ in all our models.
In general terms, the intensity of the magnetic field regulates the density contrast between the different phases.
It has a negligible effect on the ionized region, whose density is basically set by the value of $n_0$.
The density and thickness of the atomic phase may be adjusted by varying $B_0$ or $\gamma$, with larger values of either quantity yielding smaller densities and more extended regions for both the atomic and molecular layers.

One could try to obtain a good fit to the observational data by modifying $B_0$ and $\gamma$ at the same time.
A rigorous analysis would require the computation of an extensive model grid, but we have experimented with several combinations, and none of them seemed to be particularly promising.
\Referee{
As an example, we show in Figure~\ref{fig_magnetic} the density and temperature structure of a model with $B_0 = 50~\mu$G and $\gamma = 1$, consistent with observations of molecular clouds \citep{HeilesCrutcher05} and radiative MHD simulations of the expansion of magnetised \hii regions \citep{Arthur+11}.
The predictions of this model are almost indistinguishable from the standard enhanced cosmic rays scenario.
More generally, all parameter choices where}
the density was high enough to produce significant H$_2$ emission at $2.121~\mu$m led also to a very high dust extinction and thus to surface brightness profiles similar to those shown in Figure~\ref{fig_H2_line}.

However, these tests do not exhaust the available degrees of freedom.
In particular, there is no reason why the gas density and the magnetic field should be uniform across the illuminated face.
According to \citet[see their Figure~4]{Wen95}, the electron density near the ionization front ranges from $\sim 2000$ to $\sim 7000$~cm$^{-3}$ along a cut perpendicular to the Bar,
\Referee{
and observations frequently show multiple peaks in the H$_2$ emission as one moves away from the Trapezium across the Bar.
In many cases, some of these peaks occur inside the main ionization front \citep[closer to the Trapezium than the \sii peak, see e.g. Figure~4 of][]{Walmsley+00}, providing further evidence that the Bar geometry is quite complex, with multiple overlapping filaments.
}
Any realistic model of the Orion Nebula should account for these changes in the gas density encountered by different rays, which are probably associated with similar variations in the magnetic field.

\subsection{\Referee{Stellar mass distribution}}

\Referee{
In addition to the gas density at the illuminated face and the equation of state of the magnetic field, we have also considered different prescriptions to account for the gravitational acceleration associated to the stellar component.
The influence of other stars in the Trapezium has been studied by changing the central mass from 50 to 200~M$_\odot$ \citep[see e.g.][]{HerbigTerndrup86,HuffStahler06}, and we have also considered a continuous mass distribution of the form
\begin{equation}
 M(r) = M_* \frac{ r }{ r_* }
\label{eq_HH98}
\end{equation}
with $M_* = 2000$~M$_\odot$ and $r_* = 2$~pc to represent the whole stellar population of the Orion Complex \citep{HillenbrandHartmann98}.
}

\begin{figure}
\centering\includegraphics[width=8cm]{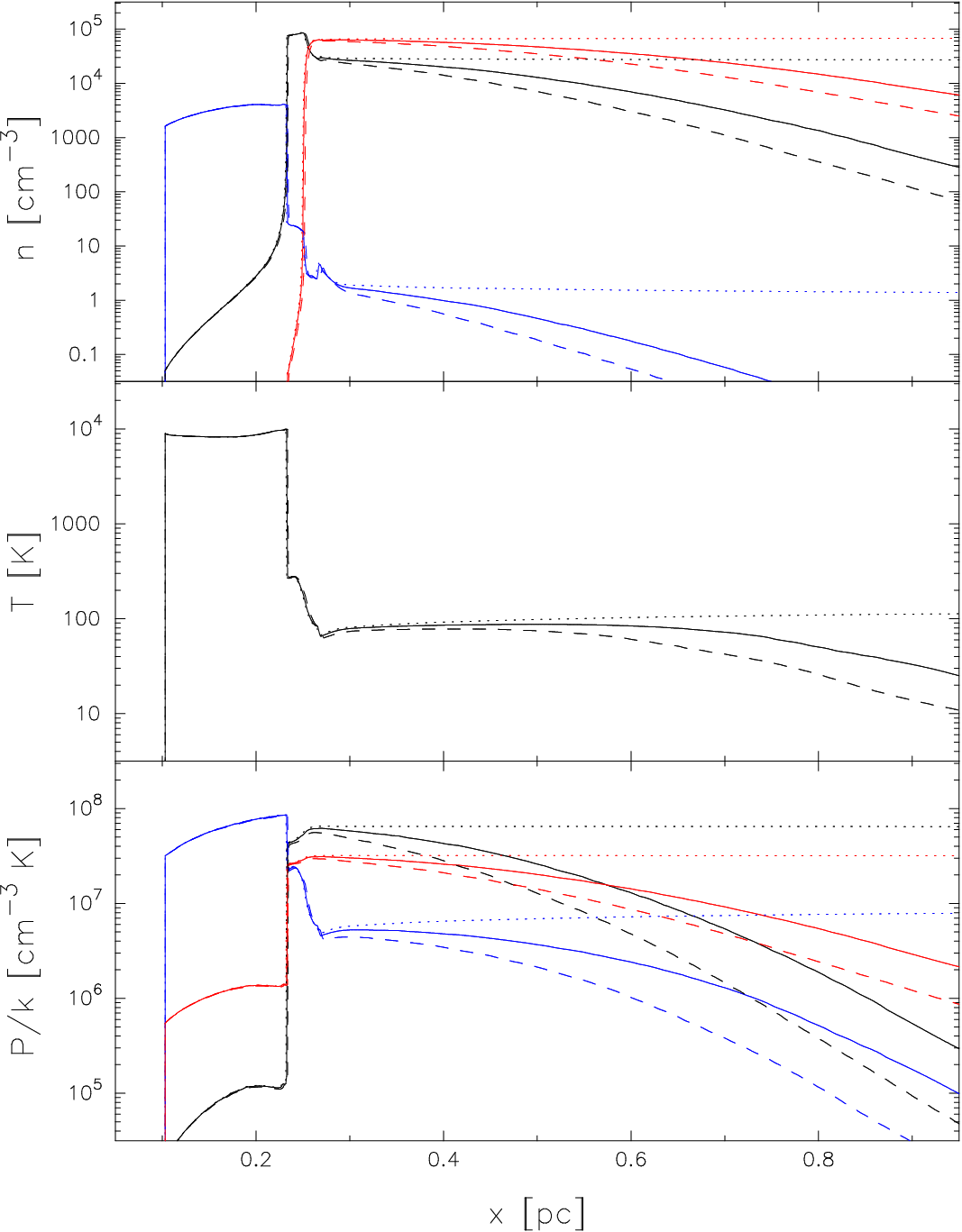}
\caption
{
\Referee{
Results obtained in the absence of gravity (dotted lines), considering \oric as a point mass of 50~M$_\odot$ (solid lines), and using equation~(\ref{eq_HH98}) to model the total stellar mass distribution (dashed lines).
Top panel: number density of ionized (blue), atomic (black), and molecular (red) hydrogen.
Middle panel: electron temperature.
Bottom panel: thermal (blue), magnetic (black), and turbulent (red) pressure components.
}
}
\label{fig_gravity}
\end{figure}

\Referee{
The effect of the gravitational acceleration is shown in Figure~\ref{fig_gravity}, where we compare the gas density, temperature, and pressure profiles predicted by the enhanced cosmic rays model in the absence of gravity, our standard implementation (self-gravity of the gas plus \oric), and a model where the stellar mass is represented by equation~(\ref{eq_HH98}).
}
As in the plane-parallel case discussed in \citet{Asc10}, the physical conditions of the gas in the ionized and neutral layers are barely affected.
In the outer parts, both density and pressure decrease with radius in order to balance gravity, and the net effect is to effectively limit the extension of the molecular cloud.
Although there is a qualitative difference with respect to the non-gravitational case, in which the molecular layer is infinite, \Referee{the prescription adopted to model the stellar distribution} does not alter the results significantly.

\Referee{
The importance of the gravitational term in equation~(\ref{eq_P}) can be assessed by comparing the total gas pressure with the value obtained in the absence of gravity.
From Figure~\ref{fig_gravity}, one sees that gravity plays a crucial role beyond $\sim0.4$~pc from the central star.
At those distances, the magnetic pressure dominates the total budget, but, due to the equation of state with $\gamma=2$, the turbulent component eventually becomes dominant at larger radii as the gas density decreases\footnote{\Referee{The ratio between magnetic and turbulent pressure would stay constant for $\gamma=1$.}}.
}

\Referee{
The escape velocity of the system, assuming the mass is distributed according to expression~(\ref{eq_HH98}), is about 2~km~s$^{-1}$, comparable to the Alfv\'en speed predicted by the models, the turbulent velocity dispersion that we assumed for the gas, and the observed velocity dispersion of the molecular gas and the stars in the Orion Nebula Cluster \citep{JonesWalker88,Tobin+09}.
The predicted extent of the molecular cloud, of the order of 1 pc radius, is also consistent with the observed size.
However, large-scale organised bulk motions (champagne flow) and smaller scale chaotic motions, both at approximately sonic speeds ($\sim 10$~km~s$^{-1}$), have been observed in the ionized gas \citep[see e.g.][]{Garcia-Diaz+08}, whereas highly supersonic motions, of the order of a few~km~s$^{-1}$, are seen in the molecular gas \citep[e.g.][]{Tobin+09}, suggesting that the Orion nebula is not a quiescent but a dynamically young object.
}

\subsection{\Referee{Model assumptions}}

First and foremost, the structure of the cloud may be completely different if the approximation of hydrostatic equilibrium is relaxed.
\Referee{
Indeed, the principal deficiency of our model is that it does not take into account gas motions and non-steady processes, which would change the emissivity of particular lines through the effects of advection and/or non-equilibrium chemistry \citep{BertoldiDraine96,StoerzerHollenbach98,Henney+05b} and modify the pressure balance.
The fact that the global model geometry is neither spherically symmetric nor strictly plane-parallel means that there is also a formal inconsistency in the pressure balance, because it is only enforced radially but not in the transverse direction (between the different radial rays).
However, this is probably a minor point in comparison with the assumption of a constant density across the illuminated face or the neglect of dynamics.
}

\Referee{
In the restricted case of steady-state motions, the pressure balance along the ray could be dealt with by adding an extra term $P_{\rm ram} = \rho v_{\rm wind}^{2}$ and adjusting the density so that the net acceleration does not vanish, but compensates for the changes in radial velocity.
Both observations and simulations \citep[e.g.][]{Mellema+06,Arthur+11} indicate velocities of the order of 10~km~s$^{-1}$ in the ionized region.
Assuming a steady mass flux $\dot{M}$ implies the relation
\begin{equation}
 v_{\rm wind} = \frac{ \dot{M} }{ 4\pi r^2 \rho }
\end{equation}
which yields typical velocities $\sim 1$~km~s$^{-1}$ in the neutral layers.
The magnitude of the effect will thus be similar to that of changing the intensity of the magnetic field or its equation of state.
Although these arguments do not necessarily hold in the general, fully time-dependent case, we think it is unlikely that the presence of gas motions qualitatively changes the results obtained by imposing strict hydrostatic equilibrium.
}

\Referee{
An independent check of the importance of non-steady effects is provided by the ratio of the sound-crossing time to the evolution timescales of the region of interest.
Only if this ratio is small will a steady-state model be justified.
The sound-crossing time
\begin{equation}
 t_{\rm s} = \frac{ \Delta r }{ c_{\rm s} }
\end{equation}
is of the order of $10^4$ years in the ionized region (typical size $\Delta r \sim 0.15$~pc and sound speed $c_{\rm s} \sim 15$~km~s$^{-1}$) and the atomic layer ($\Delta r \sim 0.015$~pc, $c_{\rm s} \sim 1.5$~km~s$^{-1}$), but it is much longer in the outer, molecular layer ($\Delta r > 0.2$~pc, $c_{\rm s} < 1$~km~s$^{-1}$).
Since these numbers are of the same order of magnitude as the evolution timescales of the corresponding layers, estimated by dividing their hydrogen column densities by the particle flux at the interfaces, the steady-state approximation seems to be only marginally valid.

On the other hand, due to the relatively high ionization parameter, the gas cooling, hydrogen recombination, and $H_2$ destruction timescales are much shorter than the sound-crossing time throughout most of the cloud, suggesting that the effects of non-equilibrium chemistry are very limited.
The only place where the chemical and dynamical timescales may be comparable is the immediate vicinity (0.01 pc) of the dissociation front, which could affect the surface brightness profile of the $H_2$ line at $2.121~\mu$m.

These simple calculations hint that the presence of gas motions would probably not modify the qualitative picture, but they must be taken into account in order to make a quantitative comparison with observational data.
}


\Referee{Regarding other model assumptions}, the role of cosmic rays on the molecular chemistry is still unclear.
Our results, most notably the abundances of CO$^+$ and SO$^+$, hint that a large population of cosmic rays, well above the Galactic background but not as large as implied by equipartition arguments, is indeed present in the Bar, but the discrepancies found for the CN, CS, SO, and SiO molecules clearly indicate that cosmic rays alone cannot provide the ultimate solution to the problem.

In addition, some atomic data (e.g. CS) are highly uncertain, and the theoretical predictions can not be more accurate than the atomic coefficients and reaction rates on which they are based.
Another complication is that some of the observations (e.g. CS and SO) assume a two-component fluid, with dense clumps embedded in a more tenuous medium, rather than a single-phase gas.
Furthermore, Cloudy calculates the cloud structure by propagating all input continua from the illuminated face of the cloud towards the outer parts, but the gas in the molecular region is also exposed to external radiation sources, such as nearby stars \Referee{\citep[see e.g. the discussion in][concerning the role of $\theta^{2}$ Ori A in the Orion Bar]{ODellHarris10}} or the extragalactic UV background.

In our view, the present discrepancies between the models and the observational measurements highlight the complexity of \hii regions and, most notably, the surrounding PDR.
Further theoretical effort is clearly required before we can claim a full understanding of the structure and chemistry of these regions, as well as the main physical mechanisms that regulate them.

 \section{Conclusions}
\label{sec_conclusions}

Thanks to its proximity and orientation, the Orion Bar offers an exquisite view of the interface between the ionized, neutral, and molecular layers of the Orion Nebula.
In the present work, we have developed self-consistent photoionization models of the Bar and compared their predictions with observational data available in the literature.
In the light of previous works that have already constrained several physical ingredients that play an important role, such as the gas density, the magnetic field, or the abundance of cosmic rays, our study is devoted to explore the effects of the geometry of the region, including also the gravitational acceleration in the pressure balance of the gas.

All the photoionization models presented here have been computed with the plasma code Cloudy, under the assumption that the Orion Bar is in approximate hydrostatic equilibrium.
In this scenario, the radius of the fully ionized region is determined by the gas density and the flux from the central star, whereas the thickness of the atomic layer is regulated by the intensity of the magnetic field.
In the outer, molecular regions, the gravitational acceleration has to be balanced by a negative pressure gradient, and the total extent of the cloud is set by the action of gravity.

Meaningful comparisons between model predictions and observational data require the assumption of an appropriate geometry for the system.
In the particular case of the Orion Bar, previous studies have modeled the region as a plane-parallel slab of thickness $h$ oriented with a small inclination angle $\beta$ with respect to the line of sight.
Including the contribution of the rest of the Orion Nebula has a significant impact on the surface brightness profiles of the emission lines produced in the ionized region, bringing them into excellent agreement with observations.

However, none of the models presented here is able to reproduce all the observable properties of the Orion Bar.
In particular, there is some mismatch in both the height and the width of the peak in the surface brightness profile of the H$_2$\,$2.121\,\mu$m emission line from the atomic region, and a severe discrepancy in the molecular column densities.
Although these results seem to be fairly robust with respect to variations in the model parameters, one possible solution would be to increase the complexity of the models -- considering, for instance, possible variations of the gas density across the illuminated face -- and use the available data to constrain the additional degrees of freedom.

 \section*{Acknowledgments}

\Referee{It is a pleasure to thank the referee, W.~Henney, for a thorough, constructive report, that helped us to improve the discussion considerably, as well as} R.~Terlevich for useful \Referee{comments on an earlier version of the manuscript}.
Funding for the present work has been provided by the Spanish \emph{Ministerio de Educaci\'on y Ciencia} (project AYA2007-67965-C03-03) \Referee{and the \emph{Comunidad Aut\'onoma de Madrid} (P2009/ESP-1496)}.

 \bibliographystyle{mn2e}
 \bibliography{references}

\appendix

\section{Cloudy input script}
\label{app_script}

Below we give an example of an input Cloudy script used to calculate the grid of photoionization models for the enhanced cosmic rays scenario.

\begin{verbatim}
c -----------------------------------------------
title - Hydrostatic model of the Orion Bar -
c -----------------------------------------------
radius 17.55 18.50
c -----------------------------------------------
c Incident continuum
c -----------------------------------------------
CMB
table read "../../star_kurucz_39600.dat"
Q(H) 49
brems 6
luminosity 32.6 range 36.77 to 588.3 Ryd
c -----------------------------------------------
c Density and chemical composition
c -----------------------------------------------
hden 3.2
abundances orion
grains pah
atom H2
c -----------------------------------------------
c Gravity
c -----------------------------------------------
constant pressure
gravity spherical
gravity external 50 Msun
c -----------------------------------------------
c Non-thermal components
c -----------------------------------------------
turbulence 2.0 km/s
magnetic field -5.1 2
cosmic ray equipartition
c -----------------------------------------------
c Stopping conditions
c -----------------------------------------------
stop temperature off
stop eden -4
set nend 10000
c -----------------------------------------------
c Output files
c -----------------------------------------------
punch overview "overview.txt" last
punch pressure "pressure.txt" last
punch molecules "molecules.txt" last
punch lines emissivity "emissivity.txt" last
S II 6716
S II 6731
H  1 6563
H2   2.121m
end of lines
punch grain extinction "extinction.txt" last
c -----------------------------------------------
c                         ... Paranoy@ Rulz! ;^D
c -----------------------------------------------
\end{verbatim}

\end{document}